\documentclass[twocolumn,superscriptaddress,amsmath,amssymb,prl]{revtex4}
\usepackage{epsfig}
\usepackage{graphics}
\usepackage{color}
\usepackage{dcolumn}
\usepackage{bm}%
\begin{document}
\title{Trapping-detrapping fluctuations in organic space-charge layers}
\author{Anna Carbone}
\email{anna.carbone@polito.it}
\affiliation{Physics Department and CNISM, \\ Politecnico di Torino, Corso Duca degli Abruzzi 24, 10129 Torino, Italy}
\author{Cecilia Pennetta}
\email{cecilia.pennetta@unisalento.it}
\affiliation{Dipartimento di Ingegneria dell'Innovazione and CNISM, \\ Universit\`{a} del Salento, 73100 Lecce, Italy}
\author{Lino Reggiani}
\email{lino.reggiani@unisalento.it}
\affiliation{Dipartimento di Ingegneria dell'Innovazione and CNISM, \\ Universit\`{a} del Salento, 73100 Lecce, Italy}


\begin{abstract}
A trapping-detrapping model is proposed for explaining the current fluctuation  behavior in organic semiconductors (polyacenes) operating under {\em current-injection conditions}. The fraction of ionized traps  obtained from the current-voltage characteristics, is related to the relative current noise spectral density at the {\em trap-filling transition}. The agreement between theory and experiments validates the model and provides an estimate of the concentration and energy level of deep traps.
\end{abstract}

\maketitle
Polymeric as well as small molecular semiconductor electronic products have shown impressive improvements in their performance during recent years.
Organic field-effect transistors (OFETs) and light-emitting diodes (OLEDs), solar cells and memories successfully compete with traditional electronic devices  when flexibility, large area, low cost and weight are the main requirements \cite{Muccini,Fleissner,Chen}. The injection of charge carriers at molecule-metal interfaces
plays a decisive role in the performance of organic
semiconductor devices. Defect states drastically alter the
injection mechanism, hence their origin and effect on charge transfer across the interface is one of the main investigation issues.
Charge carrier trapping-detrapping processes have been indeed recognized as source of current instability and degradation of organic FETs  \cite{Lang,Miyadera,Koch,Yang,Knipp,Dinelli,Schwalb,Deboer,Kang,Chandra,Giulianini}.
\par
Current and voltage noise have been widely exploited as a spectroscopic tool for investigating the role of deep defect states in inorganic insulators and wide-bang gap semiconductors operating under current-injection conditions \cite{Kleinpenning,Vanvliet,Pennetta,Verleg,Mazzetti,Tsormpatzoglou}.
A comparatively smaller number of works have been addressed to noise in organic devices  \cite{Carbone05, Jurchescu,SampietroNecliudovMartin}.
Relative current spectral densities obtained on polyacenes have evidenced a striking peaked behavior at voltages corresponding to the trap filling transition (TFT) between Ohmic ($\Omega$) and Space Charge Limited Current (SCLC) regimes.
This behavior has been interpreted as evidence for continuous percolation between the two regimes, considered as different electronic phases.
Accordingly, at the TFT, the clustering of insulating regions should lead to a reduction of the ohmic paths.
The overall effect was expected to give a substantial increase of noise in analogy with the increasing of fluctuations near a structural phase transition \cite{Carbone05}. However, a comprehensive quantitative investigation of noise behavior in organic materials is still lacking.
\par
Here, we present a complementary interpretation based on trapping-detrapping processes of injected carriers at the TFT. Accordingly, we carry out a quantitative comparison between theory and experiments, for both transport and noise.
The trap filling transition takes place in the range of voltages $ V_t \div ({c+1}) \ V_t$, with $V_t$ the threshold value for the onset of TFT and $c$ a constant whose value depends on the steepness of the transition.  These voltage values correspond to the energy range where the quasi-Fermi level crosses the trap level.
In \cite{Kleinpenning}, the relative-noise crossover between Ohmic and SCLC regimes was investigated in the absence of TFT in silicon. Here, that model is generalized by accounting for the presence of the trap-filling region between the Ohmic and SCLC regimes.
For  a two terminal sample of length $L$ and cross-sectional area $A$, the expressions of the current-voltage ($I$-$V$) characteristics and the spectral density of voltage fluctuations can be written as follows.
\par
1. \textbf{Ohmic regime}:
\begin{subequations}
\begin{eqnarray} \label{Ohm}
I_{\Omega} &=& \frac{e \mu  n_{o} A}{L}   V \hspace{10pt}, \\
S_{\Omega}(V, f) &= & \frac{\alpha_{\Omega}}{AL  n_{o} f}   V^2 \hspace{10pt},
\end{eqnarray}
\end{subequations}
where $e$ is the elementary charge, $n_o$ the density and $\mu$ the mobility of free thermal carriers, $f$ the  frequency  and $\alpha_{\Omega}$ the Ohmic Hooge parameter.
The relative noise power spectral density is:
\begin{equation} \label{relOhm}
\frac{S_{\Omega}(I, f)}{I_{\Omega}^2} = \frac{S_{\Omega}(V, f)}{V^2} = \frac{\alpha_{\Omega}}{f} \frac{1}{AL  n_{o} }\hspace{10pt}.
\end{equation}
Eq.~(\ref{relOhm}) implies  that  the relative noise is constant in the Ohmic regime as voltage increases.
\par
2. \textbf{SCLC regime}:
\begin{subequations}
\begin{eqnarray} \label{SCLC}
I_{\mathrm{SCLC}}&=& \frac{9A \epsilon_0 \epsilon_r \mu \Theta }{8 L^3}   V^2 \hspace{10pt}, \\
S_{\mathrm{SCLC}}(V,f) &=& \frac{\alpha_{\mathrm{SCLC}}}{f}\frac{ e L }{5 A \epsilon_0 \epsilon_r \Theta}   V \hspace{10pt},
\end{eqnarray}
\end{subequations}
where $\epsilon_r$ and $\epsilon_0$ are, respectively, the relative dielectric constant of the material and the vacuum permittivity, $\alpha_{\mathrm{SCLC}}$ the SCLC Hooge parameter.
The product $\mu \Theta$ appearing in Eq.~(\ref{SCLC}) is usually referred to as \emph{effective mobility} $\mu_{eff}$ with $\Theta$ given by:
\begin{equation} \label{theta2}
\Theta = \frac{n}{g n_t}=\frac{n_{v}}{g n_{t}} \exp \left[- E_{t}/k T \right] \hspace{10pt},
\end{equation}
where $n$ and $n_t$ are respectively the free and trapped carrier density, $g$ is the trap degeneracy factor, $n_v$ the density of states in the valence band (for $n$-type transport the density of states in the conduction band  $n_c$ should be considered), $E_t$ the  trap energy level, $kT$ the thermal energy.
The relative noise power spectral density is:
\begin{equation} \label{relSCLC}
\frac{S_{\mathrm{SCLC}}(I, f)}{I_{\mathrm{SCLC}}^2} = 4   \frac{S_{\mathrm{SCLC}}(V,f)}{V^2} = \frac{4 e L}{5 A \epsilon_0 \epsilon_r \Theta} \frac{\alpha_{\mathrm{SCLC}}}{f}   \frac{1}{V} \hspace{10pt}.
\end{equation}
Equation (\ref{relSCLC}) implies  that  the relative noise decreases as $1/V$  in the SCLC regime as voltage increases.
\par
\begin{figure}
 \begin{center}
  \includegraphics[width=7cm]{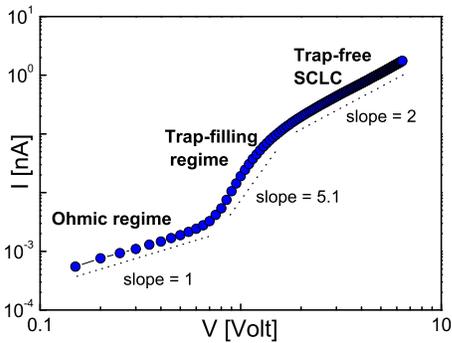}
 \end{center}
 \caption{\label{fig1} Current voltage characteristics at $T=300\mathrm{K}$ for a tetracene thin film  of
length $L= 0.65  \mu \mathrm{m}$ and cross-sectional area $A=  0.1  \mathrm{{cm}^2}$.
Circles refer to  experimental data \cite{Carbone05}.
Dot lines indicate the Ohmic region at the low voltage, the
TFT  region between $V = V_t \approx 0.8 \mathrm{V}$ and  $V \approx 2 \ V_t$ and the SCLC region at the high voltage $V > 2 \ V_t$. Solid line  is obtained by fitting the data by Eq.~(\ref{itotal}).}
\end{figure}
\begin{figure}
 \begin{center}
  \includegraphics[width=7cm]{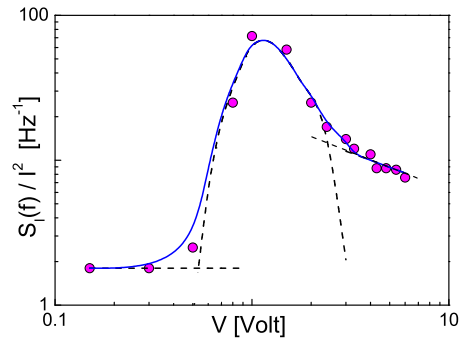}
 \end{center}
 \caption{ \label{fig2}Relative current noise spectral density $\mathcal{S}(f)$ at $f=20 \ Hz$ and T=300 K
vs applied voltage. Circles refer to experimental data \cite{Carbone05}. Dashed lines represent, respectively: (i) the Ohmic noise component at low voltages, (ii) the trapping-detrapping noise component at intermediate voltages, (iii) the SCLC noise component at high voltage. Solid line is obtained by summing the three noise components according to Eq.~(\ref{total}).}
 \end{figure}

3. \textbf{TFT regime}:\\
To the purpose of describing the noise at the TFT region, we consider the trapping-detrapping processes as the main source of fluctuations.
Following the standard approach within a linear kinetics of trapping processes \cite{Vanvliet}, the variance of the total number of ionized traps is $\overline{\Delta N^2}=N_t u(1-u)$ where $N_t$ is total number of traps in the volume of the device. Hence, the relative noise power spectrum writes:
\begin{equation} \label{TFTnoise}
\frac{S_{\mathrm{TFT}}(f)}{N_t^2} =   \frac{4}{N_t}   \frac{\tau}{1 + (2 \pi f  \tau)^2} \ u (1  - u) \equiv B \ u (1  - u) \hspace{10pt}.
\end{equation}
where $\tau$ is the carrier free time and $u=AL n_t/N_t= {1}/\{1+g \exp[(E_t-E_F)/kT]\}$ the fraction of ionized traps.
Accordingly, $u$ is a function of the applied voltage through the position of the quasi-Fermi level with respect to the trap energy level.
It is worth noting that $u(1 - u)$ vanishes at the extremes of the TFT  and reaches the maximum value at $u=0.5$, corresponding to half the filling of the total number of traps.
The total relative current noise as function of the applied voltage is then given by the sum of three independent contributions as follows:
\begin{equation}\label{total}
\mathcal{S}(f)=\frac{S_I(f)}{I^2}=
\frac{S_{\Omega}(f)}{I_{\Omega}^2} + \frac{S_{\mathrm{TFT}}(f)}{I_{\mathrm{TFT}}^2} + \frac{S_{\mathrm{SCLC}}(f)}{I_{\mathrm{SCLC}}^2}\hspace{10pt},
\end{equation}
where $I_{\mathrm{TFT}}$ is given by the Mark-Helfrich law \cite{mark62}:

\begin{equation}
\label{TFT}
 I_{\rm TFT}=A N_v\mu e^{1-l}\left[\frac{\epsilon l}{N_t(l+1)}\right]^l \left(\frac{2l+1}{l+1}\right)^{(l+1)}\frac{V^{l+1}}{L^{2l+1}}
\end{equation}
Equation~(\ref{total}) is the major assumption of the present model and is applied to the case of tetracene.
To this purpose, the function $u(V)$ is obtained by fitting the $I$-$V$ characteristics and then introduced in Eq.~(\ref{TFTnoise}) as described below.
The total current $I$ is written as sum of an Ohmic and an SCLC current generator \cite{Giulianini}.
In correspondence of the TFT regime the relative weight of the two current generators depends on the applied bias through the fraction of filled traps $u(V)$:
\begin{equation}\label{itotal}
I = I_{\Omega} + u(V) I_{\mathrm{SCLC}} \hspace{5pt}.
\end{equation}
For the validation of the model, we refer to the set of experiments carried out on polyacenes purified by
sublimation, evaporated on glass at $10^{-5}$Pa and room
temperature. Sandwich structures with Au, Al and ITO electrodes (namely Au/Pc/ITO,  Au/Tc/Al,
Au/Pc/Al) with area $A=0.1 cm^2$, distant $L=0.40\div 1.00 \mu m$ with
$\delta L=0.05 \mu m$, have been characterized. This large sample set ensures
a reliable statistics over the wide variability of chemical and structural properties. Results of $I$-$V$ and  ${S_{I}(f)}/{I^2}$ for a
tetracene thin film of length $L=0.65 \ \mu \mathrm{m}$ and cross-sectional area $A=0.1 \ \mathrm{{cm}^2}$ are shown.
From the values of the Ohmic and SCLC region of $I$-$V$ characteristics, using Eq.~(\ref{Ohm}), Eq.~(\ref{SCLC})
and $\epsilon_r =3.5$, we find $L/(Aen_{o}\mu)  = 3  \times 10^{11} \Omega$ and  $9A \epsilon_0 \epsilon_r \mu \Theta /(8 L^3) = 6 \times  10^{-11} \ A / V^2$.
From the SCLC region of the $I$-$V$ characteristics,  the effective mobility $\mu_{eff}=\Theta \mu= 4.7 \times  10^{-10} \ \mathrm{cm^2/s V}$ is obtained.
From the Ohmic region  of the $I$-$V$ characteristics,  $n_{o} \mu = 1.4 \times   10^6  \mathrm{(m V s)^{-1}}$ is obtained.
Since an independent determination of the Ohmic mobility is not available, we assume
that  $\mu$ ranges between $10^{-3} \div 1 \ \mathrm{ cm^2/Vs}$, which are reasonable values for low and high quality crystalline materials.
Accordingly, by keeping the same level of fit, the values of the parameters related to the mobility can take values in the range $\Theta \approx 10^{-7} \div 10^{-10}$ and  $n_o =10^7 \div 10^4 \ \mathrm{cm^{-3}}$.
By using Eq.~(\ref{theta2}) with  $n_{v} = 10^{21} \ cm^{-3}$, $n_{t} = 10^{13} \mathrm{\ cm^{-3}}$, it is $E_t \approx 0.85 \div 1 \ \mathrm{ eV}$.
The values of the function $u(V)$ obtained from the fit of the $I$-$V$ characteristics (see Fig. \ref{fig1}) are then introduced in Eq.~(\ref{TFTnoise}) to reproduce the noise behavior at the trap-filling transition.  Figure \ref{fig2} shows the experimental relative current spectral density (circles) and its decomposition in terms of Eq.~(\ref{total}) (dashed lines).
The quantity $B$, fitting the total excess noise in the TFT regime, is given by $B=S_{max}/0.25 = 2.8 \times 10^{-10}$~s, where $S_{max}$ is the maximum value of the relative noise.
The  maximum value of trap concentration compatible with the model used for fitting is
$n_t = 0.8 \times 10^{13} \mathrm{cm^{-3}}$.
For this value the carrier free time is  $\tau=0.8 \times 10^{-2} \mathrm{s}$.
The range of parameter values obtained from the fit are summarized in Table \ref{Table1}.
As illustrated by Figs. \ref{fig1} and \ref{fig2}, the theory reproduces quite well both transport and noise behavior at the trap-filling transition.
Furthermore, the values of the parameters are compatible with those reported in the literature \cite{Deboer}.
By taking the values of the power spectral density at $f = 20 \ Hz$, we find $\alpha_{\Omega} \approx 10^{-9} \div 10^{-12}$ and $\alpha_{\mathrm{SCLC}} \approx 10^{-7} \div 10^{-10}$. We remark that  the Hooge parameter $\alpha_{\Omega}$ differs from $\alpha_{\mathrm{SCLC}}$ to account for the difference in the tetracene medium when going from Ohmic (neutral traps) to SCLC (charged traps) conditions.
These values are low when compared with the typical values of about $10^{-3}$ in most crystalline semiconductors. These low values can be ascribed to the fact that transport takes place in insulators via a microscopic mechanism of hopping type.
\par
In conclusion, by introducing a trapping-detrapping noise contribution associated with the presence of a  TFT regime, we have developed a microscopic model which provides a quantitative interpretation of the current voltage and relative current noise in tetracene as function of the applied voltage.
The noise mechanism in the TFT region is basically related to the fluctuating occupancy  of a single level of traps, which modulates the cross-section of the conducting channels and, thus, produces noise.
The consistent fit of transport and noise provides a set of parameters (see Table  \ref{Table1}) of valuable interest for the characterization of the material.
We remark that the measured current noise spectrum  was found to be more likely to an $1/f$-sloped rather than  Lorentzian, thus, the lifetime estimate should be taken as rather indicative. The broadening of the trap level energy could account for such a spectral difference \cite{Vanvliet,Pennetta}.
Finally, we  note that the proposed model can be generalized to the case of several trap levels and hopping transport.
\par
\begin{acknowledgements}
The support of D.~Kotowski, B.~Kutrezba-Kotowska and M. Tizzoni
is gratefully acknowledged.
\end{acknowledgements}

\begin{table}
\caption{\label{Table1} Relevant parameter range for tetracene.}
\begin{ruledtabular}
\begin{tabular}{ccc}
Density of traps &	$n_{t}$ & $\approx 10^{13}  \mathrm{ cm^{-3}}$ \\
Valence band state density 	& $n_{v}$ & $\approx 10^{21}  \mathrm{ cm^{-3}}$ \\
Carrier free time &$\tau$ & $\approx 10^{-2} \ s$ \\
Zero-field hole mobility  & $\mu$ & $\approx 10^{-3} \div 1  \ \mathrm{cm^{2}/ (s V)}$ \\
Deep trap energy level  & $E_{t}$ & $\approx 0.85 \div 1 \ \mathrm{eV}$ \\
Effective mobility parameter & $\Theta$ & $\approx 10^{-7} \div10^{-10} $ \\
Thermal free carrier concentration &$n_0$ & $\approx 10^{7} \div 10^{4} \ \mathrm{ cm^{-3}}$ \\
Ohmic Hooge parameter & $\alpha_{\Omega}$ & $\approx (10^{-9} \div  10^{-12}$)\\
SCLC Hooge parameter & $\alpha_{\mathrm{SCLC}}$ & $\approx (10^{-7} \div 10^{-10})$\\
\end{tabular}
\end{ruledtabular}
\end{table}

\end{document}